\documentclass[conference]{IEEEtran}      
\usepackage{amsmath}
\usepackage{amssymb} 
\usepackage{graphics}
\usepackage{epsfig}
\usepackage{mathptmx}
\usepackage{graphicx}  
\usepackage{subcaption}
\usepackage{hyperref}

\begin{document}

\title{LayeredSense: Hierarchical Recognition of Complex Daily Activities Using Wearable Sensors}
\author{\IEEEauthorblockN{Chak Man Lam}
\IEEEauthorblockA{\textit{chakman912@gmail.com}}
}
\maketitle

\begin{abstract}

Daily activity recognition has gained prominence due to its applications in context-aware computing. Current methods primarily rely on supervised learning for detecting simple, repetitive activities. This paper introduces LayeredSense, a novel framework designed to recognize complex activities by decomposing them into smaller, easily identifiable unit patterns. Utilizing a Myo armband for data collection, our system processes inertial measurement unit (IMU) data to identify basic actions like walking, running, and jumping. These actions are then aggregated to infer more intricate activities such as playing sports or working. LayeredSense employs Gaussian Mixture Models for new pattern detection and machine learning algorithms, including Random Forests, for real-time activity recognition. Our system demonstrates high accuracy in identifying both unit patterns and complex activities, providing a scalable solution for comprehensive daily activity monitoring.
\end{abstract}

\section{INTRODUCTION}

Daily activity recognition has become a popular area of research in recent years due to its potential for context-aware computing. Among current approaches, some researchers have used smart-phone in the pocket to detect a series of basic activity in a day(walking, running, sitting etc.)[6], while some other studies rely on wearable sensors for sports activity recognition[7]. However, to our best knowledge, most of the existing work uses supervised learning to detect activity based on the training set and what's more, much of previous work focus more on simple and repeated activities. 

Therefore, with our interest to expand the scope of activity recognition from small basic activities to a wider range of more complex activities such as doing sport, working and having meal, we come up with our new architecture design for \emph{LayeredSense}. The core idea underneath the \emph{LayeredSense} is to recognize the activities in a hierarchical manner. It is relatively hard to recognize complex activities such as playing basketball or playing soccer since the signals would present random and non-periodic patterns over a relatively long duration of those activities, however those activities could be further decomposed into repeated and small-scale actions such as running, walking, jumping, shooting a ball, dribbling, kicking a ball. We call those repeated and small-scale actions \textbf{unit patterns} which are easier to be recognized with the sliding window scheme. Based on recognizing the \textbf{unit patterns}, we then further recognize the \textbf{activities} based on the time series and distribution of \textbf{unit patterns}. For example playing sports and having meal might have completely different distribution of unit patterns, even within sports, playing soccer and playing basketball could also be differentiated by the distribution of unit patterns since kicking a ball is usually much more frequent in a soccer game while jumping is more frequent in basketball game.  Based on this idea, our system \emph{LayeredSense} consists of 3 part: (1) detect new unit pattern(s) that have not been recognized in the unit pattern space. (2) recognize the unit pattern in a sliding window manner. (3) detect activities from sequence of unit patterns.\\
Our main contributions in this paper may be summarized as:

\begin{enumerate}

\item \textbf{Designing \emph{LayeredSense}, a generic architecture/framework that predict complex activities from multiple signal streams.} This system could be applied to other sensor devices and is scalable to more heterogeneous signal streams. Further, the pattern recognition algorithms adopted in this paper could be performed in both online and online real time manner, which means the system could perform training and predicting simultaneously. 

\item The system is eligible to accurately recognize and recording activities in minute resolution with adequate training, which is almost close to production level as a mobile application.

\item \textbf{Extending the clustering (GMM) model as a method to discover new unit pattern if it does not belong to existing clusters}

\item \textbf{Designing the hierarchical recognition scheme for activity recognition} Applying the bag of words algorithm to the signal processing area to recognize complex activities from unit pattern sequence.
\item Developing GUI that enable user interaction to enable supervised learning algorithm.

\end {enumerate}


\section{Signal Preprocessing}

\subsection{Myo Device}

For our project, we used an evolutionary wearable device called Myo, designed by Thalmic Labs in California.It is shown in Figure.1 below. Myo is a gesture control armband monitors people's muscle activity so that to distinguish the gestures by collecting the electromyography(EMG) data. What's more, in additional to the EMG signal, Myo is also able to monitor arm's inertia movement by collecting the inertia measurement unit. For this paper, we focus only on the imu data collected from the armband.
\begin{figure}[thpb]
      \centering
      \includegraphics[width=0.8\linewidth]{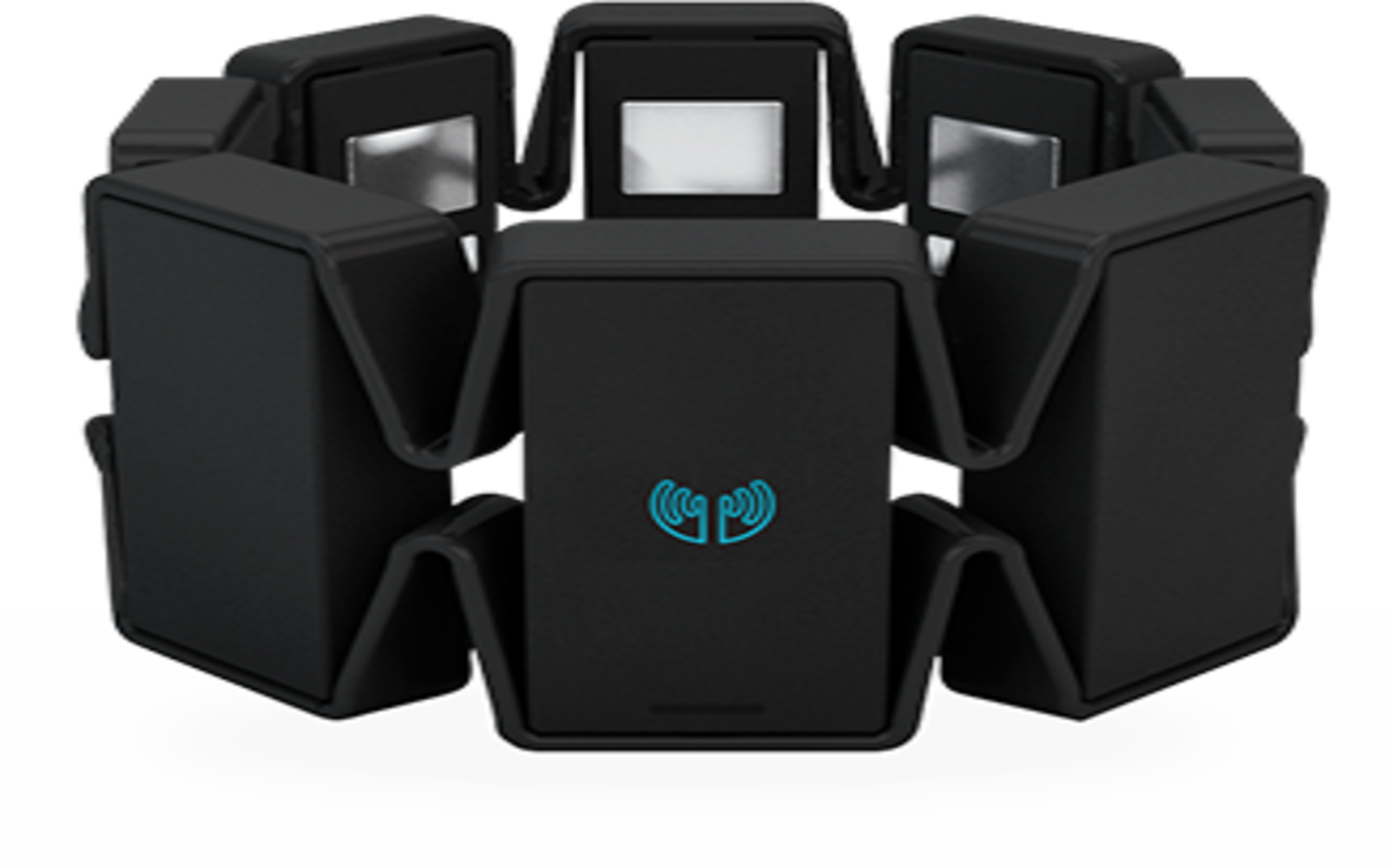} 
      \caption{Myo Armband}
      \label{img_label1}
\end{figure}

\subsection{Signal Collection and preprocessing}
As mentioned in the earlier section, the myo armband will provide two kinds of data: emg data and imu data. The emg data is a package of eight dimensional bio-electric pulse collected from the eight pieces of bio-sensors modules on the armband. The imu data is another package from the analysis of arm's movement. The imu data package is composed of nine dimensional data which includes three-axis of accelerometer, three-axis of gyroscope and three-axis of quaternion which consists of roll,pitch,yaw. Figure.2 shows the IMU signal we obtained from the Myo device for a particular shooting gestures. We took the signal in a sampling rate of 20Hz and we made a packet of signal every two second. For each packet we collected, we computed the mean, median, variance and the mean cross counts(how many times the signal passes the mean). This helps us form a 36 dimensions of features data which plays an important role in our project. We will discuss it in our later section.

\section{\emph{LayeredSense} System Overview}
In this section, we will briefly explain how our activity recognition system work. As mentioned in Section I, the main functionality of our system can be divided into three sub-blocks: new unit pattern recognition, detecting unit pattern from signal, recognizing activities from sequence of unit patterns.\\

Our experiment utilize 9 signal streams input from Myo armband include 3-axis accelerometer, 3-axis gyroscope and 3-axis roll pitch yaw data streams, each stream has a sample rate of 20 samples/second. A sample of 9 data streams is displayed in the following figure\\

In the following sub sections, we will first talk about the basic design of our framework.  We then will briefly explain how we handle the data stream from the Myo device in real time. For the rest three sub sections, we will explain the three main functionalities in detail.

\subsection{Feature generation: unit pattern}
\emph{LayeredSense} use a sliding window of 2 seconds and overlapping of 75\% (i.e. 40 samples per sliding window and step size of 10 samples).  Thus each sliding window has dimension of $40 \times 9$ (9 different signals), we calculate the statistical features, include mean, median, variance and number of mean crossing, of each of the 9 features to generate a representation of 36 dimension features. 

\subsection{New unit pattern detection}
The first main function we want to talk about is new pattern discovery. As our system is called "semi-supervised", the new pattern discovery function is the key for us to distinguish the new unit pattern from the existing training set. When our program starts, it has already includes several basic unit pattern such as walking, standing, sitting, running. We used a classifying model called Gaussian Mixture Model(GMM)[1] to partition the training data and we will explain the model later. After the program starts running, the system will analysis the overlap ratio between the existing data set and the new signal window. The overlap ratio will then compare with the probability threshold provided by the GMM. If the overlap ratio of three consecutive time windows exceeds the threshold, the system will recognize that the user may perform a movement that the system has already stored. On the other hand, if the overlap ratio of three consecutive time windows are all below the threshold, the system will consider the concurrent movement to be a new unit pattern. Then the system will pre-store the data and ask the users if they want to keep this unit pattern. If users say "yes", the system will store the unit patterns and classify the existing patterns again using GMM.\par
For the new unit pattern recognition algorithm, we have tried several existing methods. The first algorithm we tried is called Dynamic Time Warping(DTW). DTW is a very popular algorithm for measuring similarity between two temporal sequences which may vary in time or speed.[2] The way how DTW works is that for two time series who want to examine the similarity between each other, every sample in one signal will computes its distance between the sample with the same timestamp, the samples with one timestamp earlier and one timestamp later, and pick up the minimum of the three to be the distance each time. Then, start from the beginning from one time series, the two signals will compute the minimum distance(or the cost) between each sample and its neighbours, and then output a value which is cumulative cost between two time series. The lower the cost, the more similar the two signal is. However, based on our observation, the signals collected from myo device are very noisy, and also, the DTW will suffer from the scaling problem. Therefore, the DTW algorithm does not work well for matching our data streams.\par
The second method we tried is called K-means clustering, which is a classic algorithm in machine learning for computing similarity among signals based on their features. One benefit that clustering algorithm has over DTW is that it will not suffer by the scaling problem. For our system, as we initially have four basic unit patterns in the database(walking, running, standing, sitting, jumping), we computed the clusters with the number of components to be five based on the 36 features we already obtained. The clusters that are initially computed is shown in Figure.2. However, one other problem we observed is that it is hard for us to set up a common threshold for each unit pattern because each unit pattern has its own distinct variance. Therefore, instead of using absolute threshold(classify the data by value), we used relative threshold instead(classify the data by probability). The way we realize it is using another machine learning model called Gaussian Mixture Model(GMM).\par
Gaussian Mixture Model(GMM), which is similar to K-means Clustering, is also an algorithm that computes the similarity between two data streams. The difference between GMM and K-means Clustering is that GMM assume existing data is distributed according Gaussian Distribution and estimate each data's probability based on the maximum likelihood method. The equation is shown below:
\begin{equation}
P(x) = \sum_k \pi_k N(x|\mu_k,\Sigma_k)
\end{equation}
Therefore, the probability of each sample is based on the mean of each cluster and the variance of any two clusters. However, obviously the probability in the equation above cannot be solved in close-form equation[4], we used Expectation-Maximization(EM) algorithm to estimate the result. The EM algorithm is an iterative procedure for finding maximum likelihood estimates of parameters in a particular statistical model. Basically, the EM method alternatives between expectation step, which finds influence of each component on data and maximization step, which re-evaluate models using influence as weights on the given data points. After several times of iteration, the estimate probability will eventually converge, which is the final value we want. The details of how to use EM algorithm on GMM can be found in[4].\par
While we need to compute the distance between each data sample to the mean of each cluster for setting up the threshold, GMM allows us to compute the probability of each data sample in the Gaussian Model, which helps us to set a common threshold for every cluster. The clusters initially computed by GMM are shown in Figure.3. \par
The advantage of GMM over K-means Clustering is that GMM can distinguish the unit pattern that has not been detected more clearly. For instance, from the figures above we observed that while K-means clustering algorithm partitions the whole data set into 5 pieces, each GMM model forms a small cluster. If there are several points which are distributed along the border of two partitions of K-means Clustering, the system cannot tell which cluster these points should belong to, or even if the points should not belong to any cluster. However, if we apply GMM on such data points, we can obviously observed if the points are way off the clusters(new unit pattern) or they are within the existing clustering(exist unit pattern).\par
For our system, we initially applied all of the 36 features as variables of multivariate Gaussian distribution. However, as we later observed that GMM forms a very small cluster for each unit pattern and the system could not even distinguish the existed unit pattern, we decreased our features dimension to 27 where we gave up the meancrossing count of the 9 dimensional data. \par
\begin{figure}
\centering
\begin{minipage}{.5\textwidth}
	\centering
	\includegraphics[scale=0.3]{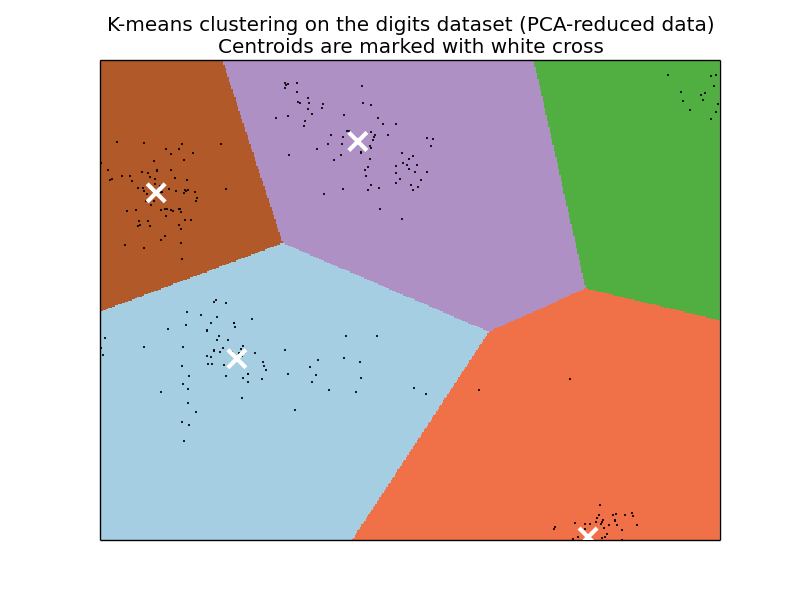}
	\caption{Pretrained data partition by K-Means Clustering}
\end{minipage}

\end{figure}

\subsection{User interaction}
As new unit pattern detected, we would add the new unit pattern to our existing patterns to increase searching space for unit pattern recognition. In the \emph{LayeredSense} system, once new unit pattern  detected, our \emph{LayeredSense} App would pop up a window to remind the user new pattern has been detected and whether she/he want to save such unit patterns, user have option to tell \emph{LayeredSense} system that it is not a pattern of interest,  ignore the new pattern, or save the new pattern. In the last case, the user is required to input the unit pattern name and corresponding activity name for labelling and training purpose. \emph{LayeredSense} requires user to repeatedly perform the unit patterns to collect the training samples, right now \emph{LayeredSense} would by default collect 120 samples until it notifying user to stop performing the unit patterns. Through the user interaction on mobile Application we developed, \emph{LayeredSense} could interactively expanding trained models to recognize more and more unit patterns and activities in a simulated supervised manner. The following figures show the user interface of processing new unit patterns and the daily activity recording.

\subsection{Unit pattern recognition}
As we generate 36 feature representation for each sliding window and the labels from user input, we apply the machine learning algorithm includes Support Vector Machine, Decision Tree and Random Forest Algorithm to recognize the unit patterns. The Random Forest achieves the best performance in accuracy and Support Vector Machine presents inferior performance. We will discuss the disadvantage of SVM in our experiment scenario in the sub section of \emph{Important observation} and result of Random Forest in the section of \emph{Performance Evaluation}

\subsection{Feature generation: unit pattern sequences}
As the \emph{LayeredSense} system recognizing the unit patterns from sliding window of signal streams, a time series of unit pattern could be generated and we try to infer the activities out from the sequence. We utilize the bag-of-words (BoW) algorithm, which is most commonly present in natural language processing literature, to infer the corresponding activities. In a typical NLP context, researchers try to predict the topic from a sentence, a paragraph or an article, which essentially is sequence of words. As an analogy to our situation, we try to predict the activity from a sequence of recognized unit patterns. \\
We use the simple version of BoW algorithm that count the number of occurrence of unit patterns in a sequence. Below is an example from our experiment.

Further we argue that the BoW is a particularly good method in our situation on 2 aspects if compared with a text recognition task:
 \begin{itemize}
 \item  are relatively stable, for example the activity of playing basketball, the distribution of unit patterns such as running, jumping, or shooting a ball is relatively stable compared different word distribution in a literature such as fiction.
 \item The unit pattern distribution of activities is less sensitive to the random positioning, the different words-ordering may affect the semantic meaning of a literature but the random ordering of unit patterns such as jumping, running, or shooting a ball does not affect the activity of playing basketball much.
 \end {itemize}

\subsection{Activity Recognition}
As \emph{LayeredSense} generate the unit pattern count as the representation features of the pattern sequence, \emph{LayeredSense} again apply the decision tree/random forest algorithm to recognize the activities.\\
In our experiment, to eliminate the noise signals we use majority vote of every 3 sequential unit patterns to display a predicted unit patterns, we set default to collect every 120 sequential patterns to recognize a predicted activities, this design result to the unit pattern prediction of every 2 seconds and the activity prediction of roughly every 60+ seconds.\\

\subsection{Important observations}
\subsubsection{Random Forest VS. Support Vector Machine} 
One important observation is that we found Decision Tree/Random Forest are better fits for real time signal stream recognition over Support Vector Machine. One typical preprocessing for the SVM algorithm is to scale the the data across the feature dimensions. However, particularly for the unit pattern recognition, \emph{LayeredSense} tests the real time stream against the SVM model trained with previous data that might has no relation in the data distribution. In concrete example, the real time stream data might have completely different mean and variance of the previous training data (as user performs completely different activities), which makes appropriate data normalization very difficult over different time period. On the other hand, the algorithmic natural of decision tree and random forest doesn't necessary requires data normalization, which is particularly good for signal data streams. Our results also indicate the decision tree/random forest greatly outperform the SVM. 
\subsubsection{Threshold in GMM model to detect new unit patterns}
Although GMM gives us a probability of each data stream, which helps us to choose a common threshold for every cluster, it is still hard to choose an unique threshold value due to the noise of signal and the different levels of similarity between each unit pattern. Therefore, for now we just choose a threshold for matching the existing pattern and a threshold for telling a new unit pattern. 

\section{Performance Evaluation}
To evaluate the \emph{LayeredSense} system, we have conducted a experiment to test three major components of the system:

\begin{enumerate}
\item accuracy of detecting new unit patterns
\item accuracy of recognizing unit patterns
\item accuracy of recognizing activities
\end {enumerate}

\subsection{Accuracy of recognizing new unit patterns}
For the evaluation of new unit pattern discovery, due to the time limitation, we have not compile this part of code into the whole real system. Therefore, we tested this method in offline. We pretrained 5 movements(reading,walking,running,shooting,dribbling), and then run the test for both existing movements and new movements. For each kind of movements, we repeated the action ten times for ten time windows each(approximately 4 seconds), computed the probability and determined if the unit pattern belongs to the existing training data set. The table below shows the percent that each unit pattern has been recognized.\par
\begin{table}[htb]
\begin{tabular}{|r|r|r|r|r|r|r|r|r|r|}
  \hline

  \hline 
  \textbf{} & \textbf{reading} &\textbf{walking} & \textbf{shooting} & \textbf{dribbling} & \textbf{running} \\
  \hline
  \hline
   reading &  50\% & 0\% & 0\% & 0\% & 0\% \\
  \hline
   walking &  0\% & 50\% & 0\% & 0\% & 0\% \\
  \hline
   shooting &  0\% & 0\% & 60\% & 0\% & 0\% \\
  \hline
   dribbling &  0\% & 0\% & 0\% & 70\% & 10\% \\
  \hline
   running &  0\% & 0\% & 0\% & 20\% & 50\% \\
  \hline
   boxing(new) &  0\% & 0\% & 0\% & 0\% & 0\% \\
   \hline
   cutting(new) &  0\% & 0\% & 0\% & 20\% & 50\% \\
  \hline
   toothing(new) &  0\% & 0\% & 0\% & 0\% & 0\% \\
  \hline
   typing(new) &  20\% & 0\% & 0\% & 0\% & 0\% \\
    \hline
    
 \end{tabular}
 \\
 \caption{Experiment results: new pattern discovery}   

\end{table}
From the table above we observed that our algorithm works well for most of the unit patterns we tested.

\subsection{accuracy of recognizing unit patterns and activities}
We design an experiment of recognition based on 9 unit patterns
and 3 activities are displayed in TABLE I \& II: \\

\begin{table}[htb]

\begin{tabular}{|r|l|}
  \hline

  \hline 
  \textbf{unit pattern labels} & \textbf{explanation} \\
  \hline
  \hline
    shooting & shooting a basketball \\
  \hline
    walking & walking \\
    \hline
    running & runing \\
   \hline
    dribbling & dribbling a basket ball \\
    \hline
    guitar\_sitting & playing Guitar while sitting  \\
    \hline
    guitar\_standing & playing Guitar while standing  \\
    \hline
    guitar\_foot\_on\_chair & playing Guitar with one foot on chair  \\
    \hline
    idle\_sitting & idle while sitting \\
    \hline
    idle\_standing & idle while standing \\
  \hline
\end{tabular}
\\
\caption{Experiment setup: Unit pattern space}

\begin{tabular}{|r|p{4.5cm}|}
  \hline

  \hline 
  \textbf{Activity labels} & \textbf{unit Pattern Combination} \\
  \hline
  \hline
    LIVE\_CONCERT & guitar\_standing,   guitar\_foot\_on\_chair, 		guitar\_sitting, running, walking  \\
  \hline
    GUITAR\_PRACTICE & guitar\_sitting, idle\_sitting \\
    \hline
    PLAY\_BASKETBALL & running, walking, shooting, dribbling \\

    \hline
    
 \end{tabular}
 \\
 \caption{Experiment setup: Activity space}   

\end{table}
For the unit pattern recognition, we collected 120 samples for each labelled unit pattern and total of $120 \times 9 = 1080 $ samples. Then we applied Random Forest algorithm for recognizing unit patterns over 75\% of randomly shuffled data and testing on the rest 25\% data. The average prediction accuracy over 4-fold cross validation is above 98\%. We then testing the trained model with real time signal streams that we perform with control, i.e. we perform actions corresponding to different unit patterns in a fixed time duration to enable us to denote the testing labels manually and test against the predicted labels from \emph{LayeredSense}, we achieve roughly 82\% accuracy of unit pattern recognition in online real time manner. Since the sliding window based prediction has temporal natural, it is reasonable to modify the measurement based on the majority vote over 3, 5 or more consecutive predicted labels as one perdition over longer period of time, the accuracy could be further improved based on majority vote schema. \\

After recognizing the unit patterns, we have a time series of predicted unit patterns, our experiment applies bag of words algorithm over every 120 unit pattern sequence to generate unit pattern count as features. We collected 20 samples for each of three activities shown in TABLE II, and applied the 4-fold cross validation similarly as above over Random Forest of activities. The average accuracy is 87\% offline and the online real time recognition accuracy is roughly the same: 83\%. One thing worth mention in our experiment control is the experimenter purposely restricts his activity to be the same in every 120 sliding windows/unit pattern period (roughly 60 seconds long) during the real time testing phase. If, however, the experimenter doesn't control the time window, say experimenter performs the unit pattern combination of \emph{PLAY\_BASKETBALL} in first 45 sliding window period and performs the unit pattern combination of \emph{LIVE\_CONCERT}, the prediction accuracy would be affected negatively. However this problem could be alleviated for those complex activities, for example if user playing basketball for 30 minutes, perform a live concert for 120 minutes or having a lunch for 20 minutes, only signal streams close to the first and last 'boundary' minute would be affected by the problem described above. The results are summarized in TABLE III   

\begin{table}[htb]
\begin{tabular}{|r|r|r|r|}
  \hline

  \hline 
  \textbf{level of actions} & \textbf{offline} &\textbf{online} & \textbf{majority vote over 5 prediction} \\
  \hline
  \hline
   unit patterns &  98\% & 82\% & 90\% \\
  \hline
    activities\_sitting & 87\% & 83\% & NA \\

    \hline
    
 \end{tabular}
 \\
 \caption{Experiment results}   

\end{table}

\section{LIMITATIONS and FUTURE WORK}
The performance measured and reported above is based on only one experimenter's result, although we have test on multiple users, we haven't formally measure and compare the results over multiple users due to the time issue. However, some noticeable limitations of our current version of \emph{LayeredSense} were found during the experiment:
\subsection{Calibration}
(1) The position of the armband on user's arm affects the accuracy in various extent, especially when the position during training phase and position during real time perdition are different, the prediction accuracy will be affected severely.\\
(2) If person 2 conduct the real time testing over the trained model from person 1's data, the accuracy would also be affected to an extend based on how conditions could be different between these 2 persons.
\subsection{Deployment to mobile device}
Although we have developed the \emph{LayeredSense} system and a mobile based user interface, we haven't successfully integrated the back-end system with front-end user interface. All the machine learning algorithms is right now computer based and requires redevelopment under mobile computing environment.
\subsection{Energy and storage efficiency consideration}
The Myo armband is eligible to continuously run in 24 hours. The mobile applications would be require to run continuously as well to perform daily activity recognition. With regard to the computational cost and storage requirement of our current version of system, there are much work worth exploring to improve the energy and storage efficiency
\subsection{Shift model training from local to cloud }
Our current system training the models locally, it actually would take very long time of training to reach the goal of recording all the daily activities. In order the make the \emph{LayeredSense} system has real application value, shifting the model training computation to cloud would be a good option such that more users could share the training data and once the model converge in training phase, later registers do not need to spent time on training most of the unit activities any more.
\subsection{Combine with indoor localization}
As indicated in [5], the combination of fingerprint of locations would further improve the indoor localization. We attempt to develop multimodel scheme that combine the real time activity recognition with indoor localization to improve the prediction accuracy in both area.

\section{Conclusion}
We come up the general idea of recognizing complex activities in an hierarchical manner, which realize the recognizing task from multiple signal streams to unit patterns and then to complex activities. We developed the \emph{LayeredSense} system accordingly which additionally incorporate the the function of discovery and training new patterns if it is not previously detected. Our system is designed to meet the signal stream features and the result meet our initial expectation. The \emph{LayeredSense} system is complete and generic that is capable to connect with other applications to exert more important values of daily activity recognition.

\bibliographystyle{IEEEtran}

\begin{thebibliography}{99}


\bibitem{c1} Wikipedia contributors.(2015,May12) Mixture Model.[Online].Available:
\url{http://en.wikipedia.org/wiki/Mixture model}

\bibitem{c2} Wikipedia contributors.(2015,April23) Dynamic Time Warping.[Online].Available: \url{http://en.wikipedia.org/wiki/Dynamic time warping}

\bibitem{c3} Wikipedia contributors.(2015,May15) Expectationmaximiza- tion algorithm.[Online].Available: \url{en.wikipedia.org/wiki/ Expectationmaximization algorithm}

\bibitem{c4} Paris Smaragdis (2014, August 25) Maching Learning for Signal Processing[Clustering]. Retrieved from \url{https://courses.engr.illinois. edu/cs598ps/ewExternalFiles/Lecture%2012%20- %20K- means, %20GMMs, %20EM.pdf}

\bibitem{c5} M. Azizyan, et al. SurroundSense: Mobile Phone Localization
via Ambience Fingerprinting. In MobiCom’09, 2009.

\bibitem{c6} Kwapisz, J. R., Weiss, G. M., and Moore, S. A. Activity
recognition using cell phone accelerometers. SIGKDD
Explor. Newsl. 12 (March 2011), 74–82.

\bibitem{c7} Ermes, M., Parkka, J., Mantyjarvi, J., Korhonen I., Detection
of daily activities and sports with wearable sensors in
controlled and uncontrolled conditions, IEEE Transactions on
Information Technology in Biomedicine 12, 1(2006), 20-26. 
\end{thebibliography}

\end{document}